\documentstyle[aps,prl,multicol,epsf]{revtex}

\begin{document}

\title{%
Asymptotically Exact Solution for Superconductivity near Ferromagnetic
Criticality
}

\author{%
Satoshi Fujimoto\thanks{E-mail address: fuji@scphys.kyoto-u.ac.jp}
}

\address{%
Department of Physics,
Kyoto University, Kyoto 606-8502, Japan
}

\date{\today}
\maketitle
\begin{abstract}%
We analyze an asymptotically exact solution for 
the transition temperature of $p$-wave superconductivity 
near ferromagnetic criticality on the basis of  
three-dimensional electron systems in which scattering processes are
dominated by exchange interactions with small momentum transfers.
Taking into account all Feynman diagrams in the gap equation,
we show that vertex corrections neglected in the conventional Eliashberg
formalism enhance the dynamical retardation effect of the pairing interaction,
and increase the superconducting transition temperature significantly, 
though they just give subleading corrections to 
properties of the normal state.
\end{abstract}


The recent discovery of ferromagnetic superconductors such as 
${\rm UGe_2}$, URhGe, and ${\rm ZrZn_2}$ has brought about
renewed interest in superconductivity (SC)
in the vicinity of ferromagnetic criticality (FC)\cite{sax,ao,pf}.
The proximity to FC implies a possible  
spin-fluctuation-mediated Cooper pairing between parallel 
spins\cite{fay,lev,mon,bla,rou,wan,kir,chu}.
For more than 20 years,
the transition temperatures $T_c$ of paramagnon-mediated SC
have been evaluated by solving
the Eliashberg equation in which vertex corrections 
are neglected\cite{fay,lev}, though it has been recognized by researchers 
that the vertex corrections are important 
in the paramagnon theory\cite{he}.
A few studies based on finite-order perturbative calculations have
revealed that a certain type of vertex correction increases $T_c$\cite{yone}.
Nevertheless, it is highly 
desirable to elucidate the role of vertex corrections
played in spin-fluctuation-mediated SC by non-perturbative
approaches.
To obtain a better physical insight into this issue, in the present paper, 
we analyze an electronic system 
with a pure repulsive interaction which 
is {\it exactly} proved to exhibit ferromagnetic critical behaviors and
also the transition to spin-triplet SC. 
The superconducting transition temperature $T_c$ is determined
by including full-order vertex corrections, and compared with
that obtained by neglecting vertex corrections. 
We will show that although vertex corrections to 
the single-particle normal self-energy 
are negligible near FC, 
$T_c$ is substantially increased by vertex corrections to the gap equation.

The model considered in the present paper is 
a three-dimensional (3D) electronic system 
in which the ferromagnetic spin exchange interaction is
dominated by forward scattering with small momentum transfers.
The model Hamiltonian is of the form,
\begin{eqnarray}
H=\sum_k E_kc^{\dagger}_{\sigma k}c_{\sigma k}
-\sum_{q}V(q)\mbox{\boldmath $S$}(q)\cdot \mbox{\boldmath $S$}(-q),
\label{ham}
\end{eqnarray}
where $\mbox{\boldmath $S$}(q)=\sum_k c^{\dagger}_{\alpha k+q}
\mbox{\boldmath $\sigma$}_{\alpha\beta}
c_{\beta k}$, $V(q)=U\Theta(q_c-\vert q_{\perp}\vert)$
with $\Theta(x)=1$ for $x \geq 0$ and 0 for $x<0$,
and $U$ is positive.
In this system, momentum transfers in the direction tangent to
the Fermi surface $q_{\perp}$ are restricted to a small range 
$\vert q_{\perp}\vert < q_c$, though momentum transfers in the direction
normal to the Fermi surface are not.  
The system can be regarded as a low-energy effective theory for
electron systems near FC.

Since, in 3D itinerant electron systems close to FC,
the Gaussian fixed point is stable\cite{he2,mil}, 
a RPA-like treatment of the ferromagnetic spin fluctuation
is valid for our system.
Moreover, because of the reason explained below, 
the spin correlators of model (\ref{ham}) are almost exactly given by
RPA-like expressions {\it with unrenormalized parameters}.
It is noted that model (\ref{ham}) is a variant of the fermion systems 
with strong forward scattering that has been
studied extensively by several authors\cite{met,kop,hal}.
In these systems, the expansion
with respect to a small parameter $q_c/(2k_F)$ enables 
us to treat electron-electron interaction
in all orders in terms of the coupling strength $U$. 
Since the low-energy properties of the systems are mainly determined by
scattering processes with small momentum transfers, 
the results obtained up to leading order in $q_c/(2k_F)$ 
are {\it asymptotically exact} at sufficiently low temperatures.
To implement this analysis, we exploit the remarkable 
theorem proved by Metzner et al. and Kopietz et al.; 
for leading order in $q_c/(2k_F)$, 
fermionic loop diagrams with more than two insertions cancel each other
in strong forward scattering fermion systems that do not involve
spin flip processes\cite{met,kop}.
The proof given by these authors for this loop cancellation theorem
can be straightforwardly generalized to 
the case of spin exchange interactions in the paramagnetic state;
i.e. {\it the loop cancellation theorem holds, even when spin flip occurs
at vertices.}
However, we should keep in mind the following limitations of the theorem.
In the ferromagnetic state,
the cancellation among loop diagrams involving spin flip is not complete,
as is easily checked by considering a loop with three legs involving
spin exchange processes. 
Thus in the following we mainly concentrate on the paramagnetic state.
A more serious restriction of our argument is that the theorem is not
applicable to the superconducting state below $T_c$, because   
loop diagrams involving the condensation of Cooper pairs 
do not cancel\cite{com}.
Despite these limitations, the theorem is still useful in analyzing
the transition temperature $T_c$, which is determined
by the linearized gap equation as shown below. 

The loop cancellation theorem ensures that the effective interaction
between electrons with parallel spins for model (\ref{ham})
is the RPA-like form,
\begin{eqnarray}
D_q(\omega_n)=\frac{-V^2(q)\chi_0(q,\omega_n)}
{1-V(q)\chi_0(q,\omega_n)} 
\approx \frac{-3U}{t+(\frac{q}{2k_F})^2+\frac{3\pi \vert\omega_n\vert}
{2vq}}.  \label{int}
\end{eqnarray}
Here $\chi_0(q,\omega_n)=-T\sum_{k,m}G^0_k(\varepsilon_m)
G^0_{k+q}(\varepsilon_m+\omega_n)$ with 
$G^0_k(\varepsilon_m)=1/(i\varepsilon_m-E_k)$, 
and $t=3[1/(UN(0))-1]$, $N(0)=k_F^3/(\pi^2E_F)$.
$v$ is the Fermi velocity.
Also, the effective interaction which involves spin exchange 
$D_q^{(ex)}(\omega)$
is given by $D_q^{(ex)}(\omega)=2D_q(\omega)$ because of the spin
rotational symmetry.
In a similar manner, we can express exact spin correlation functions as
RPA forms with unrenormalized parameters.
Thus the transition to a ferromagnetic state is completely
determined by the Stoner condition.
In the following, 
we apply our analysis only to the paramagnetic case $UN(0)<1$.

In general, RPA expressions with unrenormalized parameters are exact  
in the case where interactions between fluctuations of order parameters
are infinitely long-ranged, 
and thus all self-energy corrections vanish\cite{sha,fu}. 
We would like to emphasize that, in our system, the absence of 
self-energy corrections in eq. (\ref{int}) is not due to this reason,
but caused by loop cancellations for finite range interactions, 
and that the single-particle self-energy does not vanish by itself
in model (\ref{ham}).

\begin{figure}[h]
\centerline{\epsfxsize=8.5cm \epsfbox{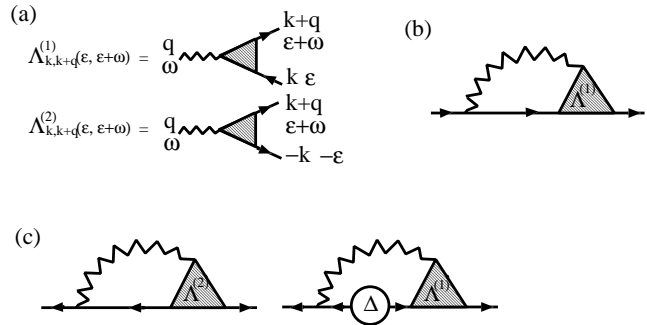}}
\caption{(a) Feynman diagrams of three-point vertices.
$\Lambda^{(1)}$ is the normal part. $\Lambda^{(2)}$ is the anomalous part.
(b) Diagram of the normal self-energy. The wavy line indicates
the effective interaction.
(c) Diagrams of the anomalous self-energy.} \label{fig}
\end{figure}

For the exact analysis of the superconducting transition, 
we calculate three-point vertices
including both normal and anomalous parts, whose diagrammatic expressions
are shown in Fig. 1(a).
As shown by Metzner et al.,
in systems with strong forward scattering, the irreducible current vertex
$\mbox{\boldmath $\Lambda$}_{k,k+q}(\varepsilon,\varepsilon+\omega)$ 
is related to the irreducible density
vertex $\Lambda_{k,k+q}(\varepsilon,\varepsilon+\omega)$ as
$\mbox{\boldmath $\Lambda$}_{k,k+q}(\varepsilon,\varepsilon+\omega)
\approx \mbox{\boldmath $v$}_{k+q/2}
\Lambda_{k,k+q}(\varepsilon,\varepsilon+\omega)$, because of
the velocity conservation\cite{met}.
Then, we obtain the asymptotic Ward identity for 
the normal part of the density three-point vertex,
\begin{eqnarray}
\Lambda^{(1)}_{k,k+q}(\varepsilon,\varepsilon+\omega)
=\frac{G^{-1}_{k+q}(\varepsilon+\omega)-G^{-1}_{k}(\varepsilon)}
{i \omega-\mbox{\boldmath $q$}\cdot\mbox{\boldmath $v$}_{k+q/2}},
\label{vert}
\end{eqnarray}
where $G_k(\varepsilon)$ is the exact single-particle Green's function.
The anomalous part of
the density three-point vertex also satisfies
the generalized asymptotic Ward identity,
\begin{eqnarray}
\Lambda^{(2)}_{k,k+q}(\varepsilon,\varepsilon+\omega)
=-\frac{\Delta_k(-\varepsilon)+\Delta_{k+q}(\varepsilon+\omega)}
{i \omega-\mbox{\boldmath $q$}\cdot\mbox{\boldmath $v$}_{k+q/2}}.
\label{verta}
\end{eqnarray}
Here $\Delta_k(\varepsilon)$ is the anomalous self-energy, which emerges
as a result of the condensation of Cooper pairs.

Using the Ward identities (\ref{vert}) and (\ref{verta}), and eq. (\ref{int}),
we can write down 
the exact normal self-energy 
$\Sigma_k(\varepsilon)$
expressed diagrammatically in Fig. 1(b), 
\begin{eqnarray}
&&\Sigma_k(\varepsilon)=
-g\sum_{k',\varepsilon'} 
D_{k-k'}(\varepsilon -\varepsilon')
G_{k'}(\varepsilon')\Lambda^{(1)}_{k,k'}(\varepsilon,\varepsilon') 
\nonumber \\
&&=
g\sum_{k',\varepsilon'}
\frac{D_{k-k'}(\varepsilon-\varepsilon')G_{k'}(\varepsilon')
G^{-1}_k(\varepsilon)}{i(\varepsilon'-\varepsilon)
-(\mbox{\boldmath $k$}'-\mbox{\boldmath $k$})\cdot
\mbox{\boldmath $v$}_{(k+k')/2}}+gC,  \\
&&C=-\sum_{k'\varepsilon'}\frac{D_{k-k'}(\varepsilon-\varepsilon')}
{i(\varepsilon'-\varepsilon)
-(\mbox{\boldmath $k$}'-\mbox{\boldmath $k$})\cdot
\mbox{\boldmath $v$}_{(k+k')/2}},
\label{self1}
\end{eqnarray}
where the prefactor $g=3$ stems from the SU(2) spin rotational symmetry, and
the exact single-particle Green's function is
$G_k(\varepsilon)=[Z_k(\varepsilon)i\varepsilon-E_k]^{-1}$
with $Z_k(\varepsilon)=1-\Sigma_k(\varepsilon)/(i\varepsilon)$.
The anomalous self-energy $\Delta_k(\varepsilon)$ at $T=T_c$ 
can be linearized in terms of the gap function 
$\Delta_k(\varepsilon)/Z_k(\varepsilon)$.
The exact linearized gap equation 
expressed diagrammatically in Fig. 1(c) is then written as,  
\begin{eqnarray}
&&\Delta_k(\varepsilon)= 
\sum_{k',\varepsilon'}G_{-k'}(-\varepsilon')
D_{k-k'}(\varepsilon-\varepsilon')
\Lambda^{(2)}_{k,k'}(\varepsilon,\varepsilon') \nonumber \\
&&+\sum_{k',\varepsilon'}\Delta_{k'}(\varepsilon')
G_{k'}(\varepsilon')G_{-k'}(-\varepsilon')D_{k-k'}(\varepsilon-\varepsilon')
\Lambda^{(1)}_{k,k'}(\varepsilon,\varepsilon')  \nonumber \\
&&=2\sum_{k',\varepsilon'}\Delta_{k'}(\varepsilon')
G_{k'}(\varepsilon')G_{-k'}(-\varepsilon')D_{k-k'}(\varepsilon-\varepsilon')
\Lambda^{(1)}_{k,k'}(\varepsilon,\varepsilon') \nonumber \\
&&+\Delta_k(\varepsilon)\sum_{k',\varepsilon'}
\frac{D_{k-k'}(\varepsilon-\varepsilon')G_k(-\varepsilon')}
{i(\varepsilon'-\varepsilon)
-(\mbox{\boldmath $k$}'-\mbox{\boldmath $k$})\cdot
\mbox{\boldmath $v$}_{(k+k')/2}} \nonumber \\
&&-\sum_{k',\varepsilon'}
\frac{\Delta_{k'}(\varepsilon')
D_{k-k'}(\varepsilon-\varepsilon')G_{k'}(\varepsilon')G_{k'}(-\varepsilon')
G^{-1}_k(\varepsilon)}{i(\varepsilon'-\varepsilon)
-(\mbox{\boldmath $k$}'-\mbox{\boldmath $k$})\cdot
\mbox{\boldmath $v$}_{(k+k')/2}}.
\label{anoself1}
\end{eqnarray}
Note that the above equations are {\it exact} in the sense that 
full-order Feynman diagrams are taken into account,   
going beyond the conventional Eliashberg equations.
In eqs. (\ref{self1}) and (\ref{anoself1}), 
the effects of superconducting fluctuations, 
which generally decrease $T_c$ and 
give rise to pseudogap phenomena\cite{ya,fu}, 
do not exist, because the loop cancellation 
suppresses their contributions. 
Therefore, $T_c$ is exactly determined by
competition between the pairing interaction in (\ref{anoself1})
and the pair-breaking effect caused by the normal self-energy (\ref{self1}).  
Since the gap equation should be solved in the vicinity of the Fermi surface
at low temperatures, $G_k^{-1}(\varepsilon)$ in the numerator of 
the last term of eq. (\ref{anoself1}) suppresses this term 
by a factor $\sim T/E_F$ compared to the other terms.
Thus we drop it in the following. 

To carry out a more explicit analysis, 
we consider an isotropic system with a spherical Fermi surface, and
linearize the dispersion relation as  $E_k=vk$.
Then, we can easily show that the constant $C$ in eq. (\ref{self1}) 
is equal to zero.
Also, in this case, $(\mbox{\boldmath $k$}'-\mbox{\boldmath $k$})\cdot
\mbox{\boldmath $v$}_{(k+k')/2}\approx
(k_r'-k_r)v+2k_Fv\sin^2(\theta/2)$, where $k_r$ is the component of momentum
normal to the Fermi surface, and
$\theta$ is the angle between
$\mbox{\boldmath $k$}'$ and $\mbox{\boldmath $k$}$.
This approximation is adopted in the denominators of the vertex functions
in eqs. (\ref{self1}) and (\ref{anoself1}).
Note that we retain the term quadratic in the transverse momentum transfers,
$\sin^2(\theta/2)$, because, as was noted by Metzner et al.\cite{met},
this term is necessary to eliminate an artificial singularity
in the vertex functions which destroys the Fermi liquid state.
In the vicinity of FC, 
since low-energy fluctuations are dominant, 
we can neglect the momentum dependence of the self-energy,
$Z_k(\varepsilon)\rightarrow Z(\varepsilon)$. 
We also expand $\Delta_k$ into the $l$-th angular momentum components 
$\Delta_l$.
Then, integrating over $E_{k'}$, 
we recast eqs. (\ref{self1}) and (\ref{anoself1}) into
\begin{eqnarray}
&&[1-Z(\varepsilon_n)]\varepsilon_n =-gT\sum_m\int^{x_c}_0 dxx\tilde{D}_0
(x,\varepsilon_n -\varepsilon_m) 
\nonumber \\
&&\times
\biggl[{\rm sign}\varepsilon_m +
\frac{i(\Sigma(\varepsilon_n)-\Sigma(\varepsilon_m))
({\rm sign}\varepsilon_m-{\rm sign}(\varepsilon_m-\varepsilon_n))}
{\vert Z(\varepsilon_m)\varepsilon_m\vert
+\vert \varepsilon_m-\varepsilon_n\vert
+2ik_Fvx^2{\rm sign}(\varepsilon_m-\varepsilon_n)}\biggr],  \label{self2}
\end{eqnarray}
\begin{eqnarray}
&&[1-(2Z(\varepsilon_n))^{-1}]\Delta_l(\varepsilon_n) 
=T\sum_m\int^{x_c}_0 dxx\tilde{D}_l
(x,\varepsilon_n -\varepsilon_m)\frac{\Delta_l(\varepsilon_m)}
{\vert Z(\varepsilon_m)\varepsilon_m\vert} \nonumber \\
&&\times\biggl[1+
\frac{i(\Sigma(\varepsilon_n)-\Sigma(\varepsilon_m))
{\rm sign}(\varepsilon_n-\varepsilon_m)}
{\vert Z(\varepsilon_m)\varepsilon_m\vert
+\vert \varepsilon_m-\varepsilon_n\vert
+2ik_Fvx^2{\rm sign}(\varepsilon_m-\varepsilon_n)}\biggr],
  \label{anoself2}
\end{eqnarray}
\begin{eqnarray}
\tilde{D}_l(x,\varepsilon-\varepsilon')
=\frac{3\pi}{2}\frac{N(0)UP_l(1-2x^2)}{S(\varepsilon-\varepsilon')
(S(\varepsilon-\varepsilon')
+\frac{\vert Z(\varepsilon')\varepsilon'\vert}{2k_Fv})},
\end{eqnarray}
where $S^2(\omega)=t+x^2+3\pi\vert\omega \vert/(4vk_Fx)$,
$P_l(x)$ is the Legendre polynomial,
and $x_c$ is a cutoff which restricts the integral over 
the transverse momentum within the range $q_{\perp}<q_c$.

We see from eq. (\ref{self2}) that the vertex corrections to the normal 
self-energy contribute only in low-energy regions, because
of the factor, 
${\rm sign}\varepsilon_m-{\rm sign}(\varepsilon_m-\varepsilon_n)$, 
yielding only subleading
corrections compared to the first term of the right-hand side of
eq. (\ref{self2}).  
Thus, the low temperature behaviors of the normal self-energy
are dominated by a one-loop diagram without vertex corrections.
To check this more explicitly, we analytically continue 
to real frequencies, and obtain
the low temperature behaviors of the normal self-energy. 
For $T<T^{*}=t^{3/2}E_F/(1.5\pi^2)$, 
the imaginary part of the self-energy shows the Fermi liquid (FL) behavior,
\begin{eqnarray}
{\rm Im}\Sigma_k^R(0)=-\frac{27k_F^3U(\pi T)^2}{32E_Ft^{3/2}}
+O((T/E_F)^2\log(t)/t).
\end{eqnarray} 
Here the second term is the subleading contribution from 
the vertex corrections.
Accordingly, the mass renormalization factor is a finite constant,
\begin{eqnarray}
\lambda=-\frac{\partial \Sigma_k^R(0)}{\partial \varepsilon}
=\frac{3k_F^3U}{\pi^2E_F}\log(\frac{1+t}{t})+O((T/E_F)^2).
\end{eqnarray}
For $T>T^{*}$, strong spin fluctuations  give rise to 
non-Fermi liquid (NFL) behaviors,
\begin{eqnarray}
{\rm Im}\Sigma_k^R(0)=-\frac{9k_F^3U}{2\pi}\frac{T}{E_F}\log(T/T^{*})
+O((T/E_F)^2), \label{ims}
\end{eqnarray}
\begin{eqnarray}
\lambda=\frac{3k_F^3U}{\pi^2E_F}\log(\frac{1.14E_F}{T})
+O((T/E_F)^2). \label{res}
\end{eqnarray}
The leading temperature dependences of the single-particle properties
coincide with those obtained by lowest order
(self-consistent or non-self-consistent) calculations\cite{don,mori}
and renormalization group calculations\cite{he2,mil}
up to constant factors.
In the NFL regime, since the enhanced single-particle damping 
of the same order as the thermal excitation energy $T$ breaks 
the quasi-particle picture, 
a perturbative expansion up to finite order in terms of
the interaction strength is insufficient for the description
of the low-energy properties.
However, the above exact results show that 
contributions from higher-order diagrams including the vertex corrections 
(the second terms of the right-hand side of eqs. (\ref{ims}) and (\ref{res}))
are negligible for model (\ref{ham}).
This observation implies that the classical theory for ferromagnetic
fluctuation developed many years ago gives a good approximation
for our system\cite{don,mori}. 

In contrast to the normal self-energy, 
the anomalous self-energy
enjoys significant contributions from the vertex corrections,
because, as seen from eq. (\ref{anoself2}), they are not
restricted to low-energy regions. 
These vertex corrections enhance 
the dynamical retardation effect of the pairing interaction.
To examine to what extent this effect affects 
the transition temperature $T_c$,
we solve the gap equation (\ref{anoself2}) numerically.
For this purpose, as is usually done, 
we recast eq. (\ref{anoself2}) into the eigenvalue problem,
$\rho(T)\Delta_l(\varepsilon_n)=\sum_m K_{mn}\Delta_l(\varepsilon_m)$.
$T_c$ is determined by the temperature at which 
$\rho(T_c)=1$. 
We consider only the $p$-wave pairing, because it gives the highest 
$T_c$ of our model.
The superconducting $T_c$ calculated as a function of the parameter
$t$, which is the measure of ferromagnetic criticality, 
is shown in Fig. 2 for a particular set of parameters.
Since our analysis is exact up to leading order in $q_c/(2k_F)=x_c$,
the errors of the computed $T_c$ are of the order $x_cT_c$.
For comparison, we also show the transition temperatures calculated 
by neglecting the vertex corrections, $T_{c0}$, in the same figure.
The asymptotically exact $T_c$ is {\it always higher than}
$T_{c0}$, even if we take into account the errors of the order $x_cT_c$. 
The maximum ratio of $T_c$ to $T_{c0}$ is approximately $\sim 3$.
It is noted that in the vicinity of the FC ($t\sim 0$),
the transition temperature decreases as $t$ approaches 
zero. This tendency is similar to the results obtained from
the conventional Eliashberg equations\cite{rou}.
According to Roussev and Millis\cite{rou}, $T_c$ does not vanish at $t=0$
within the Eliashberg formalism.
Although it is rather difficult to compute $T_c$ at $t=0$ 
with numerically reliable accuracy
from eqs. (\ref{self2}) and (\ref{anoself2}), 
we speculate that it is non-vanishing, since the vertrex corrections
always increases $T_c$.
The above results indicate that model (\ref{ham}) is 
a rare example of a purely repulsive electron system
that has been {\it exactly} proved to 
undergo the superconducting transition\cite{sha}.

\begin{figure}[h]
\centerline{\epsfxsize=7cm \epsfbox{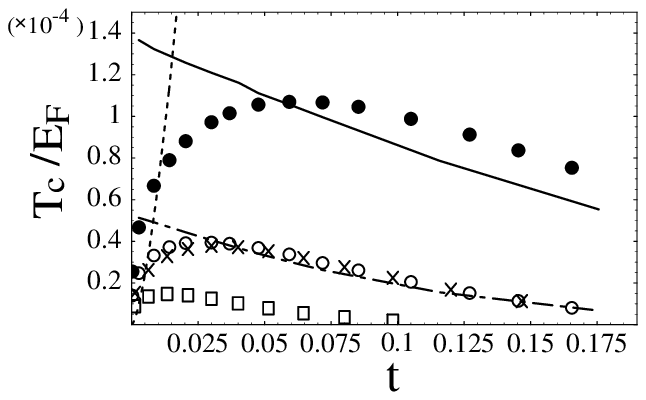}}
\caption{$T_c/E_F$ versus $t$ for $x_c=0.3$ ($\bullet$) and
$x_c=0.2$ ($\times$).
$T_{c0}/E_F$ versus $t$ for
$x_c=0.3$ ($\circ$) and $x_c=0.2$ ($\Box$). 
The broken line indicates the crossover temperature $T^{*}$ which distinguishes
the NFL state (high temperature region) and 
the FL state (low temperature region).
The solid and dashed-broken lines indicate 
$T_c \times 0.1/E_F$ and $T_{c0} \times 0.1/E_F$ for $x_c=0.3$
in the Ising case, respectively. } \label{fig2}
\end{figure}

Next we consider the case with the Ising-like anisotropy
of the exchange interaction.
To simplify the calculation, we assume that 
the Ising anisotropy of spin degrees of freedom
is so strong that the interaction part of the Hamiltonian (\ref{ham})
is replaced with $-\sum_{q}V(q)S^z(q)S^z(-q)$.
In this case, the asymptotically exact $T_c$ is obtained 
from eqs. (\ref{self2}) and (\ref{anoself2}) 
with only one modification that the factor $g$ in (\ref{self2})
is replaced with $g=1$.
The transition temperatures computed for this Ising case are shown in
Fig. 2 (solid line). 
The transition temperature without vertex corrections, $T_{c0}$,
are also shown in the figure (dashed-broken line).
$T_c$ takes a maximum value at the ferromagnetic
critical point $t=0$ as in the case without vertex 
corrections\cite{mon,rou,wan}.
This is due to the absence of the transverse spin fluctuation
which causes the pair-breaking effect, and suppresses
$T_c$ for small $t$ in the SU(2) symmetric case.
We would like to note that in the case with the strong Ising anisotropy, 
the loop cancellation theorem holds even in the ferromagnetic metallic state.
We can calculate the asymptotically exact $T_c$ in the ferromagnetic state
by using eqs. (\ref{self1}) and (\ref{anoself1}) with $g=1$,
which may be relevant to ferromagnetic superconductors such as
UGe$_2$, URhGe, and ZrZn$_2$.

In conclusion, we have obtained the asymptotically exact solution
for the transition temperature of $p$-wave SC near FC.
We have shown that vertex corrections neglected in the conventional
Eliashberg equation enhance the dynamical 
retardation effect of the pairing interaction,
increasing the transition temperature significantly. 

The author would like to thank K. Yamada and H. Ikeda for valuable
discussions.
Numerical computations were partly performed at the Yukawa
Institute Computer Facility.
This work was supported by a Grant-in-Aid from the Ministry
of Education, Science, Sports and Culture, Japan.


\begin{thebibliography}{9}
\bibitem{com} Because of this limitation, 
our analysis is not applicable to the first order transition
to SC discussed in ref. 11).
\end{thebibliography}
\end{document}